\pdfoutput=1
\documentclass[11pt,a4paper]{article}

\usepackage[utf8]{inputenc}
\usepackage[T1]{fontenc}
\usepackage{times}
\usepackage{amsmath,amssymb}
\usepackage{graphicx}
\usepackage{booktabs}
\usepackage{hyperref}
\usepackage{url}
\usepackage{xcolor}
\usepackage{listings}
\usepackage[margin=1in]{geometry}
\usepackage{caption}
\usepackage{enumitem}
\usepackage{algorithm}
\usepackage{algpseudocode}
\usepackage{tikz}
\usetikzlibrary{arrows.meta,positioning,fit,backgrounds,calc}
\usepackage{xspace}
\usepackage{natbib}
\hypersetup{
    colorlinks=true,
    linkcolor=blue,
    citecolor=blue,
    urlcolor=blue
}

\newcommand{\ours}{\textsc{ManagerWorker}\xspace}

\title{Can AI Models Direct Each Other?\\Organizational Structure as a Probe into Training Limitations}

\author{
Rui Liu \\
Independent Researcher \\
\texttt{ult.rui.liu@gmail.com}
}

\date{}

\begin{document}

\maketitle

\begin{abstract}
Can an expensive AI model effectively direct a cheap one to solve software engineering tasks? We study this question by introducing \ours{}, a two-agent pipeline where an expensive ``manager'' model (text-only, no code execution) analyzes issues, dispatches exploration tasks, and reviews implementations, while a cheap ``worker'' model (with full repo access) executes code changes. We evaluate on 200 instances from SWE-bench Lite across five configurations that vary the manager-worker relationship, pipeline complexity, and model pairing. Our findings reveal both the promise and the limits of multi-agent direction: (1) a strong manager directing a weak worker (62\%) matches a strong single agent (60\%) at a fraction of the strong-model token usage, showing that expensive reasoning can substitute for expensive execution; (2) a weak manager directing a weak worker (42\%) performs \emph{worse} than the weak agent alone (44\%), demonstrating that the directing relationship requires a genuine capability gap---structure without substance is pure overhead; (3) the manager's value lies in \emph{directing}, not merely \emph{reviewing}---a minimal review-only loop adds just 2pp over the baseline, while structured exploration and planning add 11pp, showing that active direction is what makes the capability gap productive; and (4) these behaviors trace to a single root cause: current models are trained as monolithic agents, and splitting them into director/worker roles fights their training distribution. The pipeline succeeds by designing around this mismatch---keeping each model close to its trained mode (text generation for the manager, tool use for the worker) and externalizing organizational structure to code. This diagnosis points to concrete training gaps: delegation, scoped execution, and mode switching are skills absent from current training data.
\end{abstract}

\section{Introduction}
\label{sec:intro}

Why do companies have managers?

This question, usually confined to organizational theory, has a surprisingly direct analog in AI agent design. A software company could, in principle, staff entirely with senior individual contributors (ICs)---skilled engineers who each independently analyze requirements, design solutions, write code, and review their own work. Some small teams do operate this way. But as organizations scale, they universally adopt a hierarchical structure: managers who \emph{plan and review} coordinate ICs who \emph{execute}. This is not mere bureaucracy. The IC/manager structure persists because it provides three benefits:

\begin{enumerate}[nosep]
    \item \textbf{Asymmetric cost allocation.} Senior judgment is expensive and scarce; junior execution is cheaper and more available. The manager's time is spent on high-leverage decisions (what to build, where to look, whether the approach is correct), while the IC's time is spent on lower-leverage but necessary execution (reading files, writing code, running tests).
    \item \textbf{Iterative oversight.} The manager does not fire-and-forget. They review intermediate results, catch misunderstandings early, and redirect effort before it compounds. This feedback loop is more efficient than having the IC self-correct, because the manager has broader context and can spot systematic errors.
    \item \textbf{Separation of concerns.} The manager thinks at the architectural level (which files, what approach, what risks) without being distracted by implementation details. The IC focuses on making the code work without worrying about whether they are solving the right problem. This division prevents the ``tunnel vision'' that occurs when a single agent both plans and executes.
\end{enumerate}

Current AI coding agents are all ICs. Whether using SWE-agent~\citep{yang2024sweagent}, OpenHands~\citep{wang2025openhands}, Claude Code, or similar systems, a single model receives a GitHub issue and autonomously explores the repository, plans a fix, writes code, and iterates until tests pass. The entire cognitive budget---from root cause analysis to code editing---is spent within a single model invocation (or a single agentic session). When multi-agent systems are used, they typically treat agents as \emph{tools}---a router selects which agent is best suited for a task, or agents are composed in a fixed pipeline where each performs a specialized function. The question is always ``which agent should do this task?''

We ask a different question: \textbf{how should agents be organized?} Rather than treating agents as interchangeable tools to be selected or routed, we treat them as \emph{people in an organization} with defined roles, reporting structures, and communication protocols. The question shifts from capability matching to organizational design---not which agent is best, but what \emph{relationship} between agents produces the best outcomes. This is the difference between staffing a project (tool view) and designing a team (organizational view). In \ours{}, an expensive ``manager'' model serves as the manager: it reads the issue, decides what needs to be explored, interprets exploration results, writes a plan, reviews the worker's implementation, and provides corrective feedback. A cheap ``worker'' model serves as the IC: it has full access to the repository and executes specific tasks---exploring files, implementing patches, running commands---under the manager's guidance.

Critically, the manager is \textbf{text-only}: it never accesses the repository directly. This is both a cost optimization (text-only calls are cheaper than agentic sessions) and a design choice informed by our finding that managers who can read files tend to hallucinate ``the fix is already applied'' rather than actually analyzing the problem. The worker, conversely, has full agentic access but receives tightly scoped instructions.

The pipeline operates in iterative loops that mirror the human IC/manager workflow:

\begin{enumerate}[nosep]
    \item \textbf{Analysis:} Manager reads the issue and generates 2--3 exploration tasks (``find the \texttt{from\_file} method in \texttt{config.py}'', ``check how \texttt{\_\_init\_\_} handles defaults'').
    \item \textbf{Exploration loop:} Workers execute tasks in parallel, report findings. Manager reviews reports, decides if more exploration is needed (up to 3 rounds).
    \item \textbf{Planning:} Manager synthesizes findings into a specific implementation plan with exact code changes.
    \item \textbf{Implementation loop:} Worker implements the plan. If the patch is invalid, manager reviews the failure and provides revised guidance. If the patch is valid, manager reviews for correctness and either approves or requests revisions (up to 3 rounds).
\end{enumerate}

We evaluate \ours{} on 200 instances from SWE-bench Lite~\citep{jimenez2024swebench} across five configurations. Our experiments address three research questions:

\begin{itemize}[nosep]
    \item \textbf{RQ1 (Direction):} Can an expensive model effectively direct a cheap one, and what capability gap is required?
    \item \textbf{RQ2 (Cost--quality tradeoff):} How does the quality--cost tradeoff curve look across different levels of pipeline complexity---from no director, to a minimal review loop, to a full structured pipeline?
    \item \textbf{RQ3 (Model limitations):} What does the multi-agent structure reveal about the limitations of current language models---as directors, as workers, and as autonomous agents?
\end{itemize}

Our key contributions are:
\begin{enumerate}[nosep]
    \item The \ours{} pipeline: a multi-agent system that applies organizational IC/manager structure to software engineering tasks, with iterative oversight loops for both exploration and implementation.
    \item A quality--cost Pareto analysis across five configurations on 200 SWE-bench Lite instances, showing that a strong manager with a weak worker (62\%) matches a strong single agent (60\%) at a fraction of the strong-model token usage, while a weak manager with a weak worker (42\%) performs worse than the weak agent alone (44\% on the same subset).
    \item An ablation via a minimal 50-line ``simple loop'' (53\%) that isolates the value of the director's presence from the value of pipeline complexity, showing that most of the gain comes from structured exploration and planning rather than mere oversight.
    \item A diagnosis of \emph{why} the structure works: current models are trained as monolithic agents, and the pipeline succeeds by keeping each model close to its training distribution---the manager generates text, the worker calls tools---while externalizing organizational structure to code. This explains both the successes (62\% with structured phases) and the failures (53\% when structure is implicit, 42\% when the manager lacks capability).
\end{enumerate}

A preview of our central finding: the IC/manager pattern works for AI agents, but for \emph{different reasons} than it works for humans. In human organizations, the structure provides oversight and judgment. In AI systems, its primary value is \emph{constraining each model to operate close to its training distribution}---preventing the manager from falling back on tool use, preventing the worker from reasoning beyond its scope, and providing phase transitions that neither model can maintain on its own. The organizational metaphor is useful for \emph{designing} the pipeline, but the explanation for \emph{why} it works is about training distributions, not about oversight.

\section{Related Work}
\label{sec:related}

\paragraph{Multi-agent systems for software engineering.}
ChatDev~\citep{qian2024chatdev} simulates an entire software company with CEO, CTO, programmer, and tester agents communicating through structured dialogues. MetaGPT~\citep{hong2024metagpt} encodes Standard Operating Procedures as multi-agent workflows. MapCoder~\citep{islam2024mapcoder} uses four specialized agents for competitive programming. HyperAgent~\citep{phung2025hyperagent} uses four specialized agents (Planner, Navigator, Code Editor, Executor) to manage the full lifecycle of SE tasks. These systems focus on \emph{role specialization}---each agent has a different function. Our approach instead focuses on \emph{capability asymmetry}---the manager and worker have different cost/capability profiles, and the structure is designed to exploit this gap. We show that when the gap is absent, the structure hurts performance.

\paragraph{Hierarchical and orchestrated agents.}
AgentVerse~\citep{chen2024agentverse} provides a framework for multi-agent collaboration with configurable group dynamics. AutoGen~\citep{wu2024autogen} enables multi-agent conversations with customizable interaction patterns. Masters et~al.~\citep{masters2025manageragent} formalize the Manager Agent as a research challenge for orchestrating human-AI teams, modeling workflow management as a Partially Observable Stochastic Game. Khattab et~al.~\citep{khattab2025scaling} derive scaling laws for multi-agent systems and find that heterogeneous capability mixing---pairing low-capability orchestrators with high-capability subagents---can outperform homogeneous systems by 31\% on certain benchmarks. Our work is closest to these orchestration approaches, but differs in a critical design choice: the manager is \textbf{text-only} (no code execution), and we pair a \emph{high}-capability manager with a \emph{low}-capability worker, the opposite direction from Khattab et~al., showing this asymmetry is essential for software engineering tasks.

\paragraph{Coding agents for SWE-bench.}
\textbf{Agentic approaches} give a single LLM autonomy to explore and edit: SWE-agent~\citep{yang2024sweagent} provides an agent-computer interface, OpenHands~\citep{wang2025openhands} offers an open platform (ICLR 2025), and AutoCodeRover~\citep{zhang2024autocoderover} combines LLMs with AST-based search. Live-SWE-agent~\citep{yang2025livesweagent} self-evolves its scaffold at runtime, achieving 77.4\% on SWE-bench Verified. \textbf{Agentless approaches} use fixed pipelines: Agentless~\citep{xia2024agentless} uses hierarchical localization followed by repair. Our pipeline is hybrid: the manager operates in a structured (agentless) manner, while the worker operates agentically within bounded tasks. Unlike Live-SWE-agent which pushes single-agent capabilities to the frontier, we show that organizational structure can substitute for model capability.

\paragraph{Planning, self-reflection, and cost-aware agents.}
Self-Planning~\citep{jiang2023selfplanning} showed that explicit planning before coding improves Pass@1 by 25.4\%. PairCoder~\citep{zhang2024paircoder} separates planning (Navigator) from implementation (Driver). SWE-Search~\citep{antoniades2025swesearch} uses Monte Carlo Tree Search for backtracking during implementation. BudgetMLAgent~\citep{gandhi2024budgetml} uses LLM cascades, pairing a free base model with occasional expert calls for planning. Our approach differs from all of these in that planning and execution are performed by \emph{different models at different cost points}, and the manager never sees code directly---it operates purely on text reports from the worker.

\section{Method: The \ours{} Pipeline}
\label{sec:method}

\ours{} assigns two roles with asymmetric capabilities:

\begin{itemize}[nosep]
    \item \textbf{Manager} (expensive model, text-only): Reads the issue description, interprets worker reports, writes plans, reviews patches. No repository access.
    \item \textbf{Worker} (cheap model, full access): Explores files, implements code changes, runs git commands. Follows manager instructions.
\end{itemize}

\begin{figure*}[t]
\centering
\begin{tikzpicture}[
    >=Stealth,
    box/.style={draw, rounded corners=3pt, minimum height=0.8cm, minimum width=2.0cm,
                font=\small, align=center, thick},
    mbox/.style={box, fill=blue!10},
    wbox/.style={box, fill=orange!12},
    lbl/.style={font=\scriptsize\itshape, text=gray!60!black},
    arr/.style={->, thick},
    darr/.style={->, thick, dashed, gray!50},
]

\begin{scope}[on background layer]
  \fill[blue!3] (-5.5, 1.5) rectangle (9.8, -1.6);
  \fill[orange!3] (-5.5, -1.6) rectangle (9.8, -4.8);
\end{scope}

\node[font=\small\bfseries, text=blue!60!black, anchor=north west] at (-5.3, 1.4) {Manager};
\node[font=\scriptsize, text=blue!50!black, anchor=north west] at (-5.3, 1.0) {(text-only, no repo)};
\node[font=\small\bfseries, text=orange!60!black, anchor=north west] at (-5.3, -1.7) {Worker};
\node[font=\scriptsize, text=orange!50!black, anchor=north west] at (-5.3, -2.1) {(full repo access)};

\draw[gray!25, thick] (-5.5, -1.6) -- (9.8, -1.6);

\node[mbox] (analyze) at (-2.0, 0) {Analyze\\[-2pt]{\scriptsize issue}};

\node[mbox] (review1) at (1.2, 0) {Review\\[-2pt]{\scriptsize reports}};
\node[wbox] (explore) at (1.2, -3.2) {Explore\\[-2pt]{\scriptsize codebase}};

\node[mbox] (plan) at (4.2, 0) {Write\\[-2pt]{\scriptsize plan}};

\node[mbox] (review2) at (6.8, 0) {Review\\[-2pt]{\scriptsize patch}};
\node[wbox] (impl) at (6.8, -3.2) {Implement\\[-2pt]{\scriptsize fix}};

\node[box, fill=green!12, minimum width=1.4cm] (out) at (9.0, -1.6) {\small Patch};

\draw[arr] (analyze.south) -- ++(0,-1.0) -| (explore.north)
    node[lbl, pos=0.15, right] {tasks};

\draw[arr] (explore) -- (review1)
    node[lbl, midway, right, xshift=2pt] {reports};

\draw[darr] (review1.west) -- ++(-0.6, 0) |- node[lbl, pos=0.25, left] {more?} (explore.west);

\draw[arr] (review1) -- (plan)
    node[lbl, midway, above] {enough};

\draw[arr] (plan.south) -- ++(0,-1.0) -| (impl.north)
    node[lbl, pos=0.15, right] {plan};

\draw[arr] (impl) -- (review2)
    node[lbl, midway, right, xshift=2pt] {patch};

\draw[darr] (review2.west) -- ++(-0.6, 0) |- node[lbl, pos=0.25, left] {revise} (impl.west);

\draw[arr] (review2.east) -- (out.north west)
    node[lbl, midway, above] {approve};

\draw[rounded corners=6pt, dashed, gray!40, thick]
    ($(explore.south west)+(-0.9,-0.3)$) rectangle ($(review1.north east)+(0.3,0.3)$)
    node[lbl, font=\scriptsize, below] at ($(explore.south)+(0,-0.3)$) {Exploration loop (max 3)};
\draw[rounded corners=6pt, dashed, gray!40, thick]
    ($(impl.south west)+(-0.9,-0.3)$) rectangle ($(review2.north east)+(0.3,0.3)$)
    node[lbl, font=\scriptsize, below] at ($(impl.south)+(0,-0.3)$) {Implementation loop (max 3)};

\end{tikzpicture}
\caption{The \ours{} pipeline. The manager (top lane, text-only) analyzes, reviews, and plans. Workers (bottom lane, full repo access) explore and implement. Loop arrows on the left side show iterative rounds. Communication is structured natural language.}
\label{fig:pipeline}
\end{figure*}
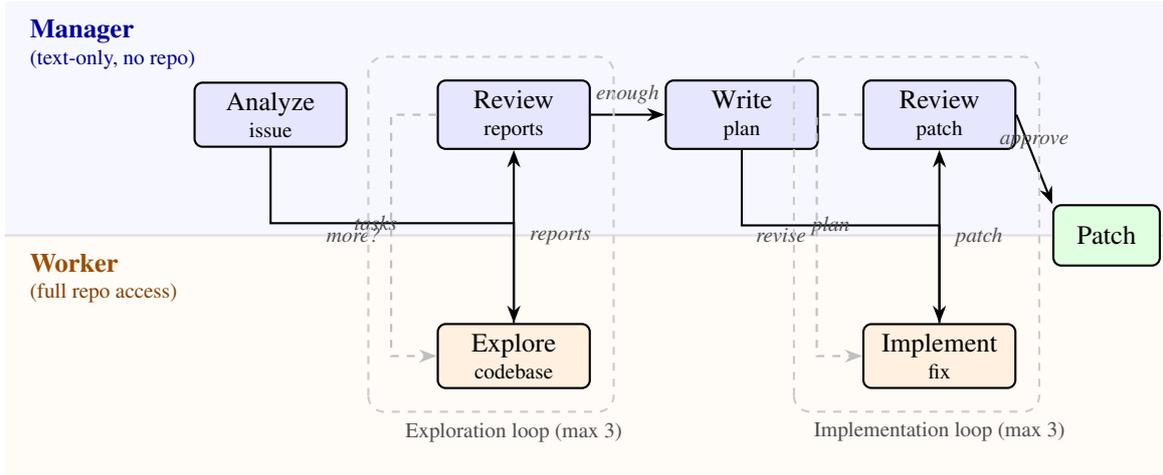

The pipeline has four phases organized into two iterative loops (Figure~\ref{fig:pipeline}).

The full procedure is given below.

\begin{algorithm}[H]
\caption{The \ours{} Pipeline}
\label{alg:pipeline}
\begin{algorithmic}[1]
\Require Issue $I$, Repository $R$, Manager $M_{\text{mgr}}$, Worker $M_{\text{wkr}}$
\Ensure Patch $P$
\Statex
\Statex \textbf{Phase 1: Analysis}
\State $\textit{tasks} \leftarrow M_{\text{mgr}}.\textsc{Analyze}(I)$ \Comment{Generate $\leq 3$ exploration tasks}
\Statex
\Statex \textbf{Phase 2--3: Iterative Exploration} \Comment{Max 3 rounds}
\State $\textit{reports} \leftarrow \emptyset$
\Repeat
    \State $\textit{reports} \leftarrow \textit{reports}\ \cup$ \Call{WorkersExplore}{\textit{tasks}, $R$} \Comment{Parallel execution}
    \State $\textit{plan}, \textit{tasks} \leftarrow M_{\text{mgr}}.\textsc{PlanOrExplore}(I, \textit{reports})$
\Until{$\textit{plan} \neq \texttt{None}$ \textbf{or} max rounds reached}
\Statex
\Statex \textbf{Phase 4--5: Iterative Implementation} \Comment{Max 3 rounds}
\State $\textit{prompt} \leftarrow \textsc{Guided}(I, \textit{plan})$ \Comment{Round 1: worker has autonomy}
\Repeat
    \State $P \leftarrow M_{\text{wkr}}.\textsc{Implement}(\textit{prompt}, R)$
    \State $\textit{verdict} \leftarrow M_{\text{mgr}}.\textsc{Review}(\textit{plan}, P)$
    \State $\textit{prompt} \leftarrow \textsc{Strict}(\textit{verdict})$ \Comment{Retries: follow corrections exactly}
\Until{$\textit{verdict} = \textsc{Approve}$ \textbf{or} max rounds reached}
\State \Return $P$
\end{algorithmic}
\end{algorithm}

\subsection{Phase 1: Analysis}

The manager receives the issue description and generates 2--3 focused exploration tasks. Each task is a natural language instruction scoped to a specific investigation (e.g., ``Find the \texttt{get\_default} method in \texttt{django/db/models/fields.py} and report its signature and the logic around line 800''). Tasks are capped at 3 to control cost; we found that generating more tasks yields diminishing returns and increases the risk of irrelevant exploration.

The analysis prompt instructs the manager to identify:
\begin{itemize}[nosep]
    \item Which files and code regions are likely involved
    \item What specific information the worker should gather
    \item What the root cause might be (as a hypothesis to test)
\end{itemize}

\subsection{Phase 2--3: Iterative Exploration}

Workers execute exploration tasks in parallel, each producing a natural language report summarizing what they found. The manager then reviews all reports and decides:

\begin{itemize}[nosep]
    \item \textbf{More exploration needed:} Generate 1--3 additional tasks based on what was learned. This happens when the initial exploration reveals unexpected code structure or when the root cause hypothesis needs revision.
    \item \textbf{Enough information:} Produce a detailed implementation plan with specific code changes.
\end{itemize}

This loop runs for at most $R_{\text{explore}} = 3$ rounds. On the final round, the manager must produce a plan regardless of confidence. The plan includes:
\begin{itemize}[nosep]
    \item Root cause analysis with supporting evidence from exploration reports
    \item Specific files and locations to modify
    \item Exact code changes (ideally as find/replace blocks)
    \item What \emph{not} to change
\end{itemize}

\subsection{Phase 4--5: Iterative Implementation}

The implementation phase mirrors the exploration loop but with different prompt strategies per round:

\paragraph{Round 1 --- Guided autonomy.} The worker receives the full issue description, the manager's plan, and instructions to ``use your judgment on the exact implementation.'' The worker is encouraged to read relevant files before editing. This gives the worker freedom to adapt the plan to the actual code structure, which may differ from what the manager assumed.

\paragraph{Round 2+ --- Strict with corrective feedback.} If the worker's patch is invalid or rejected by the manager, the manager analyzes the failure and provides revised, more specific instructions. The worker now receives these as strict instructions (``follow these EXACTLY''), because the manager has seen what went wrong and can provide precise corrections.

This asymmetric prompt strategy---guided on first attempt, strict on retries---emerged from our observation that:
\begin{itemize}[nosep]
    \item Pure strict prompts (all rounds) prevent the worker from adapting to code it discovers, causing failures when the manager's assumptions about code structure are wrong.
    \item Pure guided prompts (all rounds) allow the worker to drift from the plan, causing regressions on instances where the manager's plan was correct.
    \item Guided-then-strict combines the benefits: the worker's first attempt reveals ground truth about the code, and the manager uses this to write precise corrections.
\end{itemize}

The implementation loop runs for at most $R_{\text{impl}} = 3$ rounds. The manager can issue two verdicts on a valid patch:
\begin{itemize}[nosep]
    \item \textbf{APPROVE:} Patch looks correct, return it.
    \item \textbf{REVISE:} Patch has issues, provide specific feedback for the next round.
\end{itemize}

If no valid patch is produced after all rounds, the pipeline returns the best partial result or an empty patch.

\subsection{Design Decisions}

Several design choices emerged from iterative experimentation:

\paragraph{Manager must be text-only.} Early experiments gave the manager repository access, which consistently degraded performance (see Section~\ref{sec:analysis} for ablation data). Removing file access forces the manager to reason abstractly and delegate investigation to workers.

\paragraph{Exploration tasks are capped at 3.} In one early run, the manager generated 317 exploration tasks for a single issue, crashing the system. Even without such pathological cases, more than 3--4 tasks per round yield diminishing returns: the additional information rarely changes the plan, but the cost scales linearly.

\paragraph{Worker reports, not raw output.} Workers produce natural language summaries rather than dumping raw file contents. This compression is essential: the manager's context window cannot absorb full file contents from multiple exploration tasks, and the summarization step forces the worker to extract the relevant information.

\subsection{The Simple Loop Baseline}
\label{sec:simple_loop}

To isolate the value of the full pipeline's structure from the mere presence of a manager, we also evaluate a minimal variant: the \textbf{Simple Loop}. In this $\sim$50-line pipeline, the worker executes the task freely (with full repo access), then the manager reviews the worker's output (diffs and tool-call summaries) and either approves or provides feedback for the next round. There is no structured exploration phase, no explicit planning phase, and no guided-then-strict prompt strategy---just execute, review, repeat.

The Simple Loop tests whether a manager who merely \emph{reviews} worker output can match a manager who actively \emph{directs} exploration and implementation. As Section~\ref{sec:results} shows, the answer is no: the simple loop achieves 53\% versus the full pipeline's 62\%, confirming that the structured phases provide substantial value beyond basic oversight.

\subsection{Separating Reasoning from Execution}

The IC/manager structure decomposes a coding task into two activities: \textbf{reasoning} (analysis, planning, review---text-in, text-out, no tools) and \textbf{execution} (file reading, code editing, running commands---tool-intensive, less analytical). In a single-agent system, these are entangled in one growing context window at one cost point. \ours{} separates them, matching each to the appropriate model tier. Section~\ref{sec:analysis} examines why this separation works---and why it sometimes doesn't.

\section{Experimental Setup}
\label{sec:experiments}

\subsection{Benchmark}

We evaluate on a 200-instance subset of \textbf{SWE-bench Lite}~\citep{jimenez2024swebench}, sampled from 300 real-world GitHub issues across Django, Flask, Matplotlib, Scikit-learn, Sphinx, Sympy, and other Python repositories. Each task requires generating a patch (git diff) that resolves the issue, evaluated by running the project's test suite in a Docker container.

\subsection{Models}

We use two model tiers to test the capability asymmetry hypothesis:

\begin{itemize}[nosep]
    \item \textbf{Strong model:} Claude Sonnet 4.6 (via Copilot CLI or Claude Code SDK). Used as manager in the asymmetric configuration and as the single-agent baseline.
    \item \textbf{Weak model:} GPT-5-mini (via Copilot CLI, free tier). Used as worker in all manager configurations and as the weak single-agent baseline.
\end{itemize}

Both models are accessed through the GitHub Copilot CLI, which provides a consistent agentic interface with file access, command execution, and git operations.

\subsection{Configurations}

We compare five configurations (Table~\ref{tab:configs}):

\begin{table}[h]
\centering
\small
\begin{tabular}{@{}llccc@{}}
\toprule
\textbf{Configuration} & \textbf{Description} & \textbf{Manager} & \textbf{Worker} & \textbf{Pipeline} \\
\midrule
\ours{} & Full structured pipeline & Sonnet 4.6 & GPT-5-mini & 5-phase \\
Simple Loop & Minimal review loop & Sonnet 4.6 & GPT-5-mini & execute$\to$review \\
Strong Direct & Single strong agent & --- & Sonnet 4.6 & none \\
Weak Direct & Single weak agent & --- & GPT-5-mini & none \\
Weak$\to$Weak & Weak manager, weak worker & GPT-5-mini & GPT-5-mini & 5-phase \\
\bottomrule
\end{tabular}
\caption{Experimental configurations. Direct configurations are single-agent baselines. \ours{} and Simple Loop test two levels of pipeline complexity with the same model pairing. Weak$\to$Weak tests whether structure helps without a capability gap.}
\label{tab:configs}
\end{table}

\ours{} uses a 5-phase pipeline (analyze $\to$ explore $\to$ plan $\to$ implement $\to$ review) with iterative loops. The Simple Loop is a minimal $\sim$50-line pipeline where the worker executes freely, the manager reviews and provides feedback, and the loop repeats until approval or a round cap. It isolates the value of the manager's \emph{presence} from the value of \emph{structured} exploration and planning.

In practice, Strong Direct uses one full agentic session per instance ($\sim$30K strong-model tokens). \ours{} uses 3--7 lightweight text-only manager calls ($\sim$6.6K strong-model tokens) plus 4--12 worker agentic sessions ($\sim$60K weak-model tokens). The Simple Loop uses 2--3 manager review calls ($\sim$3K strong-model tokens) plus 2--3 worker sessions. All models are accessed through the GitHub Copilot CLI or Claude Code CLI.

\subsection{Metrics}

\begin{itemize}[nosep]
    \item \textbf{Resolve rate:} Percentage of tasks where all FAIL\_TO\_PASS and PASS\_TO\_PASS tests pass.
    \item \textbf{Empty patch rate:} Percentage of tasks where the agent produced no valid patch (indicates pipeline failures).
    \item \textbf{Evaluation error rate:} Percentage of tasks where the generated patch caused the evaluation harness to crash (indicates malformed output).
\end{itemize}

\section{Results}
\label{sec:results}

\subsection{Main Results}

Table~\ref{tab:main} presents the main results on 200 SWE-bench Lite instances.

\begin{table}[h]
\centering
\small
\begin{tabular}{@{}lccccc@{}}
\toprule
\textbf{Configuration} & \textbf{Manager} & \textbf{Worker} & \textbf{$n$} & \textbf{Resolved} & \textbf{Rate} \\
\midrule
\ours{} (full pipeline) & Sonnet 4.6 & GPT-5-mini & 200 & \textbf{124/200} & \textbf{62\%} \\
Strong Direct & --- & Sonnet 4.6 & 200 & 120/200 & 60\% \\
Simple Loop & Sonnet 4.6 & GPT-5-mini & 200 & 106/199 & 53\% \\
Weak Direct & --- & GPT-5-mini & 200 & 101/200 & 51\% \\
\midrule
\multicolumn{6}{@{}l}{\textit{50-instance subset (for Weak$\to$Weak comparison):}} \\
\ours{} (full pipeline) & Sonnet 4.6 & GPT-5-mini & 50 & 32/50 & 64\% \\
Weak Direct & --- & GPT-5-mini & 50 & 22/50 & 44\% \\
Weak$\to$Weak & GPT-5-mini & GPT-5-mini & 50 & 21/50 & 42\% \\
\bottomrule
\end{tabular}
\caption{Main results. Top: 200-instance evaluation. \ours{} matches Strong Direct (+2pp) using primarily cheap worker tokens. Bottom: 50-instance subset including Weak$\to$Weak, which performs worse than Weak Direct (42\% vs 44\%)---structure without a capability gap is net negative.}
\label{tab:main}
\end{table}

\paragraph{Finding 1: An expensive manager can substitute for an expensive worker.} \ours{} resolves 124/200 (62\%), comparable to Strong Direct (60\%) which uses the expensive model for everything. The worker (GPT-5-mini) resolves only 51\% when operating alone---the manager's guidance accounts for an 11 percentage point improvement, lifting a free-tier model to match a premium single agent. The strong model's value is in \emph{reasoning} (analysis, planning, review), not in \emph{execution} (file reading, code editing, running commands).

\paragraph{Finding 2: Structure without capability gap hurts.} Weak$\to$Weak resolves only 42\%, worse than Weak Direct. On the same 50-instance subset where Weak$\to$Weak was measured, Weak Direct achieves 44\%---the structure costs 2 percentage points. Adding organizational structure when the manager lacks sufficient analytical capability introduces coordination overhead without compensating benefits. The weak manager generates plausible-sounding but incorrect plans, and the worker faithfully wastes implementation rounds on wrong guidance. The structure \emph{amplifies} the weak model's reasoning failures rather than compensating for them.

\paragraph{Finding 3: Pipeline complexity matters, but the director matters more.} The Simple Loop---where the worker executes freely and the manager only reviews---achieves 53\%, just 2pp above Weak Direct. The full pipeline adds another 9pp by structuring the interaction into exploration, planning, and guided-then-strict implementation. This suggests that the manager's value comes primarily from \emph{directing} (telling the worker what to investigate and how to fix), not merely from \emph{reviewing} (approving or rejecting worker output).

\subsection{Quality--Cost Tradeoff}

Table~\ref{tab:cost} compares the cost profiles across all configurations. The five configurations form a Pareto frontier: increasing pipeline complexity and model capability yields diminishing returns in quality but monotonically increasing cost.

\begin{table}[h]
\centering
\small
\begin{tabular}{@{}lcccc@{}}
\toprule
\textbf{Configuration} & \textbf{Resolve} & \textbf{Strong tokens} & \textbf{Weak tokens} & \textbf{Total tokens} \\
\midrule
\ours{} (full) & 62\% & $\sim$6.6K & $\sim$60K & $\sim$67K \\
Strong Direct & 60\% & $\sim$30K & 0 & $\sim$30K \\
Simple Loop & 53\% & $\sim$3K & $\sim$30K & $\sim$33K \\
Weak Direct & 51\% & 0 & $\sim$15K & $\sim$15K \\
Weak$\to$Weak & 42\% & 0 & $\sim$75K & $\sim$75K \\
\bottomrule
\end{tabular}
\caption{Quality--cost tradeoff across configurations. Token estimates per instance based on typical prompt sizes and session lengths. Resolve rates for top four configs from $n=200$; Weak$\to$Weak from $n=50$. \ours{} achieves comparable quality to Strong Direct while shifting $\sim$90\% of tokens to the cheap model. Weak$\to$Weak is the worst of both worlds: high total tokens and low quality.}
\label{tab:cost}
\end{table}

Several patterns emerge from this tradeoff curve:

\paragraph{Strong-model tokens are the scarce resource.} \ours{} uses $\sim$6.6K strong-model tokens per instance versus $\sim$30K for Strong Direct---a $\sim$5$\times$ reduction---while achieving comparable quality (62\% vs 60\%). The manager's text-only calls are inherently cheap: no tool use, no growing context, short input and output. The expensive work (file reading, code editing, retrying) is delegated to the free-tier worker.

\paragraph{More pipeline complexity has diminishing returns.} The jump from no director (Weak Direct, 51\%) to a minimal review loop (Simple Loop, 53\%) is only +2pp. The jump from the simple loop to the full pipeline (\ours{}, 62\%) is +9pp. But the full pipeline also uses $\sim$2$\times$ the total tokens of the simple loop. Whether the additional complexity is worthwhile depends on the cost of the worker model and the value of each percentage point of resolve rate.

\paragraph{Structure without capability amplifies cost without quality.} Weak$\to$Weak uses the most total tokens ($\sim$75K) yet achieves the worst quality (42\%). The weak manager generates plans that lead the worker down wrong paths, wasting rounds of execution. This is strictly dominated by Weak Direct, which uses $\sim$15K tokens for 44\% resolve rate on the same subset.

\subsection{Degradation on Harder Instances}

To understand how the approach scales with task difficulty, we examine the Simple Loop's performance across batches of the 200-instance evaluation (Table~\ref{tab:batches}). Instances are ordered by the SWE-bench Lite canonical ordering; later batches tend to contain harder instances. We note that minor pipeline improvements were made between batches (code context in reviews, relaxed execution constraints), so batches 50--150 benefited from both easier instances and improved prompts. The final batch (150--200) uses the same pipeline as batch 100--150, isolating the effect of instance difficulty.

\begin{table}[h]
\centering
\small
\begin{tabular}{@{}lccp{5.5cm}@{}}
\toprule
\textbf{Batch} & \textbf{Resolved} & \textbf{Rate} & \textbf{Notes} \\
\midrule
0--50 & 26/49 & 53\% & Initial pipeline \\
50--100 & 30/51 & 59\% & + code context in reviews \\
100--150 & 33/52 & 63\% & + relaxed execution constraints \\
150--200 & 17/47 & 36\% & Same pipeline as 100--150; harder instances \\
\bottomrule
\end{tabular}
\caption{Simple Loop (v2) performance by batch. Batches 50--150 benefited from both pipeline improvements and easier instances, so these rates are confounded. The clean comparison is batch 100--150 (63\%) vs 150--200 (36\%): same pipeline, harder instances. This 27pp drop exposes the ceiling of the review-only approach.}
\label{tab:batches}
\end{table}

The full pipeline shows more stability: 64\% at $n=50$, 64\% at $n=100$, and 62\% at $n=200$. The structured exploration and planning phases help the manager maintain effectiveness on harder instances where the simple review loop degrades. However, the 150--200 batch remains the hardest for all configurations, suggesting fundamental limits to what direction alone can achieve.

\subsection{Complementarity with Single-Agent Approaches}

On a 50-instance subset where both \ours{} and Strong Direct were evaluated:

\begin{table}[h]
\centering
\small
\begin{tabular}{@{}lc@{}}
\toprule
\textbf{Combination} & \textbf{Resolved} \\
\midrule
\ours{} & 32/50 (64\%) \\
Strong Direct & 29/50 (58\%) \\
Union & 33/50 (66\%) \\
Intersection & 28/50 (56\%) \\
\midrule
\ours{} only & 4/50 (8\%) \\
Strong Direct only & 1/50 (2\%) \\
\bottomrule
\end{tabular}
\caption{Complementarity analysis on 50-instance subset. \ours{} resolves nearly all instances that Strong Direct resolves (28/29), plus 4 additional. The approaches solve largely overlapping but not identical subsets.}
\label{tab:complementarity}
\end{table}

The high overlap (28/29 of Strong Direct's solutions are also solved by \ours{}) suggests that the manager-worker pipeline subsumes most of the single agent's capability. The 4 instances uniquely solved by \ours{} tend to require systematic multi-file exploration, where the manager's structured analysis prevents the worker from getting lost. The 1 instance uniquely solved by the single agent involved deep iterative debugging where autonomous exploration outperformed top-down direction.

\section{Analysis}
\label{sec:analysis}

\subsection{The Core Issue: Models Are Trained as Monolithic Agents}

Current language models are trained as unified agents---they reason \emph{and} execute within a single session. The training distribution consists of conversations where the model reads files, edits code, runs commands, and reasons about results, all in one continuous context. The manager-worker split asks models to do something they were not trained for: operate in \emph{half} of this loop. The manager must reason without acting. The worker must act without reasoning strategically. Both roles fight the model's training distribution, and the specific ways in which they fight it explain our results.

\subsubsection{Directors Fight Their Training to Delegate}

When given repository access, the manager does what it was trained to do: read files and form conclusions. But this trained behavior is counterproductive in the director role. Table~\ref{tab:director_ablation} shows the effect on a 5-instance development set.

\begin{table}[h]
\centering
\small
\begin{tabular}{@{}lcl@{}}
\toprule
\textbf{Manager Configuration} & \textbf{Resolved (toy-5)} & \textbf{Key Change} \\
\midrule
v5--v6: Manager with repo access & 0--1/5 & Manager hallucinates ``already fixed'' \\
v7: Manager text-only & 2/5 & Forced abstract reasoning \\
v9: Text-only + iterative exploration & 3/5 & Manager can request more info \\
\bottomrule
\end{tabular}
\caption{Director design ablation on a 5-instance development set. Giving the manager repository access caused a regression---the model reverted to its training behavior (read files, form conclusions) rather than delegating investigation.}
\label{tab:director_ablation}
\end{table}

With repo access, the manager reads a file, matches it against its parametric knowledge, and declares ``the fix is already applied.'' This is not a reasoning failure per se---it is the model doing exactly what it was trained to do: process evidence and reach a conclusion. The problem is that the director role requires a different behavior: formulate questions, dispatch investigation, and synthesize reports from others. The text-only constraint works because it \emph{removes the model's ability to fall back on its trained behavior}, forcing it into the unfamiliar but more productive mode of abstract reasoning and delegation.

In our extended testing with a director-first variant (where the director drives every step), we observed the same pattern from the other direction: even when explicitly told ``you cannot read files, workers do that for you,'' the director model would address a human user (``Could you approve read access?''), describe fixes instead of dispatching workers to make them, or hallucinate code state from training data. The model drifts back to its training distribution---being a helpful unified agent---under pressure.

\subsubsection{Workers Fight Their Training to Follow Instructions}

The worker faces the mirror problem. When given a task-level instruction like ``investigate how \texttt{RenameField} interacts with \texttt{ForeignKey to\_fields},'' cheap models (GPT-5-mini) often \emph{narrate their intent} rather than executing and reporting:

\begin{quote}
\small\itshape
``I plan to first read the migration operations file to understand how RenameField is implemented, then check the autodetector for how it handles field references\ldots''
\end{quote}

\noindent The model produces a plan instead of results. This is its trained behavior: when given a vague instruction, generate a helpful response. But the worker role requires a different behavior: execute tool calls and return data. We found that GPT-5-mini is only reliable when given one specific, atomic instruction---it cannot autonomously decompose a task into tool calls and report the results.

The communication strategy ablation (Table~\ref{tab:iterative}) reveals a subtler version of the same problem across multiple rounds.

\begin{table}[h]
\centering
\small
\begin{tabular}{@{}lcccp{4.5cm}@{}}
\toprule
\textbf{Strategy} & \textbf{Resolved} & \textbf{Empty} & \textbf{Eval Err} & \textbf{Description} \\
\midrule
Strict only & 26/50 (52\%) & 8 & 6 & Worker follows plan exactly; no autonomy \\
Guided only & 27/50 (54\%) & 5 & 1 & Worker has full autonomy on every round \\
Guided-then-strict & \textbf{32/50 (64\%)} & 0 & 2 & Autonomy on round 1, strict on retries \\
\bottomrule
\end{tabular}
\caption{Communication strategy ablation on 50-instance subset. The guided-then-strict strategy produces the largest improvement (+10pp over strict-only), showing that cheap models need different levels of constraint at different stages.}
\label{tab:iterative}
\end{table}

Under \textbf{strict-only} prompts, the worker cannot adapt when the manager's assumptions about code structure are wrong---it fails silently and produces empty patches (8/50). Under \textbf{guided-only} prompts, the worker reverts to its trained behavior as a unified agent: it ``improves'' code beyond scope, introduces unrelated changes, or abandons the plan entirely. The \textbf{guided-then-strict} strategy works because it externally regulates the worker's mode---round 1 allows the trained explore-and-edit behavior (useful for discovering ground truth), while retries constrain the worker to follow precise corrections (preventing drift). The model cannot regulate this transition itself; the pipeline must impose it.

\subsubsection{Directing Is an Untrained Skill}

The Weak$\to$Weak result (42\% vs 44\% for Weak Direct on the same 50-instance subset) reveals that \emph{directing}---reasoning about natural language summaries of code to produce actionable plans---is a distinct skill that weaker models have not acquired.

The weak manager generates plans that are syntactically plausible but semantically wrong: correct-looking file paths that don't exist, reasonable-sounding code changes that miss the root cause, and confident assessments that misinterpret exploration reports. The worker faithfully implements these wrong plans, and the iterative loop compounds the error: each round produces a new incorrect approach rather than converging.

This is not simply ``less analytical capability.'' Writing a task brief that a worker can follow---specifying what to fix, where, and why, at the right level of abstraction---is a cognitive mode that differs from code generation. It resembles writing a technical specification or a code review comment: the author must reason about code without seeing it, anticipate implementation challenges, and communicate intent precisely enough for someone else to act on it. Models trained primarily on code completion and conversational tool use may not develop this skill, regardless of their parameter count.

Our extended testing confirmed this: even strong models (Sonnet, DeepSeek) struggle to maintain semantic-level delegation. When asked to write task descriptions for workers, they drift toward operation-level instructions (``read file X lines Y--Z'') or vague task-level instructions (``fix the bug'')---neither of which is effective. The sweet spot---semantic-level delegation that shares accumulated context and intent while leaving execution to the worker---requires a mode of communication that is largely absent from current training data.

\subsubsection{The Planning Gap}

The manager's plan is bounded by what exploration reports reveal. In our v5 experiments, instance \texttt{astropy-14182} was resolved only when the worker independently added a \texttt{start\_line} parameter that was not in the manager's plan. The worker discovered this requirement during implementation, but the manager---working from exploration reports that didn't mention this detail---could not have planned for it.

This is a structural consequence of the role separation: the manager reasons about a \emph{summary} of the codebase, not the codebase itself. Iterative exploration partially addresses this (the manager can request more information), but the gap persists for details the manager doesn't know are relevant---unknown unknowns that only emerge during implementation.

\subsubsection{Non-Determinism}

Identical configurations produce different results across runs. Versions v5e and v5f used the same code, same models, same instances---yet resolved 5/10 and 4/10 respectively. The lost instance was resolved in one run by the worker's independent discovery of a missing parameter, and missed in the other. This stochasticity is inherent to language model inference and means that small-sample results (including our 50-instance ablations) should be interpreted with caution.

\subsection{Why the Pipeline Works Despite These Limitations}

Given that both the director and worker roles fight model training, why does \ours{} work at all? Because \textbf{the pipeline design works \emph{around} the training distribution rather than against it}, constraining each model to operate in a mode close enough to its training:

\begin{itemize}[nosep]
    \item The manager generates \emph{text}---analysis, plans, review comments---which is what LLMs naturally produce. It is never asked to delegate tool calls or coordinate workers in real time, only to write structured natural language.
    \item The worker executes \emph{tool calls in a scoped context}---which is what agentic models are trained to do. It receives a clear task with instructions and works autonomously within that scope.
    \item The \emph{pipeline code} provides the organizational structure---phase transitions, round limits, prompt strategy switching---that neither model can maintain on its own. The structure is in the code, not in the models.
\end{itemize}

This explains why the full structured pipeline (62\%) substantially outperforms the simple loop (53\%): the simple loop asks the director to implicitly manage phase transitions (when to stop exploring, when to commit to a fix), while the full pipeline makes these transitions explicit in code. The more structure is externalized from the model to the pipeline, the less the model needs to operate outside its training distribution.

\section{Discussion}
\label{sec:discussion}

\subsection{The Training Distribution Problem}

Our analysis identifies a specific, actionable problem: \textbf{current models are trained as monolithic agents, and multi-agent structures require skills absent from this training distribution.}

The monolithic agent training loop---receive task, reason, call tools, iterate---produces models that are good at being individual contributors. But the manager-worker split requires two distinct skills that this training does not develop:

\begin{enumerate}[nosep]
    \item \textbf{Delegation}: reasoning about a problem and producing a task description for someone else to execute, at the right level of abstraction (not too vague, not too granular). This resembles writing technical specifications, code review comments, or engineering briefs---a mode of communication largely absent from tool-use training data.
    \item \textbf{Scoped execution}: following someone else's instructions without reasoning about whether the task is correct, staying within scope, and reporting results without editorializing. This is the opposite of the autonomous exploration that agentic training encourages.
\end{enumerate}

Our pipeline works not by overcoming these deficits but by \emph{designing around them}. The manager writes text (close to its training distribution); the worker calls tools in a scoped context (close to its training distribution); and the pipeline code provides the organizational structure that neither model can maintain. This design-around strategy has diminishing returns: the more the task requires genuine delegation or genuine scoped execution, the more the models revert to their trained behavior and the pipeline degrades.

\subsection{The Limits of Direction}

This framing clarifies where and why the directing approach breaks down:

\textbf{Hard instances require deep autonomous exploration.} The Simple Loop's batch breakdown (Table~\ref{tab:batches}) shows a sharp drop from 63\% to 36\% on the hardest instances. These bugs require many hypothesis-test cycles---exactly the kind of iterative reasoning that monolithic agents are trained for. The manager cannot anticipate these cycles from text reports, and the pipeline's structured phases are too rigid to accommodate them.

\textbf{The planning gap is a consequence of role separation.} The manager can only plan based on what exploration reports contain. When the fix requires details the manager didn't know to ask about, the plan has gaps. This is not a model limitation per se, but a structural cost of separating reasoning from execution: the manager trades direct observation for the benefits of abstraction, and sometimes the trade is unfavorable.

\textbf{Coordination overhead dominates on simple tasks.} For straightforward single-file bugs, the exploration and planning phases are pure overhead. The weak direct agent (51\%) solves some of these faster because it can immediately start editing---using exactly the monolithic agent behavior that training optimized for.

\subsection{Implications for Model Training}

Our findings suggest concrete training objectives that would improve multi-agent systems:

\begin{itemize}[nosep]
    \item \textbf{Train on delegation tasks.} Current training data consists of tool-use conversations where the model acts. Delegation requires a different kind of output: task briefs, specifications, and review comments that \emph{describe} work for others. Training on manager-worker transcripts, code review threads, and technical specification documents could develop this skill.
    \item \textbf{Train on scoped execution.} Workers need to follow instructions, stay within scope, and report results cleanly---suppressing the urge to reason about the larger problem. Training on pairs of (instruction, execution trace) where the model is rewarded for staying on task could address the drift problem.
    \item \textbf{Train for mode switching.} The guided-then-strict finding shows that the optimal behavior changes across phases: explore freely in round 1, follow instructions precisely in retries. Models that can recognize and adapt to these different modes---without external prompt engineering---would reduce the pipeline's dependence on handcrafted phase transitions.
    \item \textbf{Calibrate tool-use behavior.} The manager hallucination finding shows that models cannot be trusted to use tools judiciously. When given file access, the manager confirms rather than investigates. Training that rewards \emph{withholding} conclusions when evidence is insufficient---and delegating investigation instead---could address this.
\end{itemize}

\subsection{Limitations}

\paragraph{Single benchmark.} SWE-bench Lite contains only Python repositories. The IC/manager structure may be more or less effective for other languages, frameworks, or task types.

\paragraph{Two model tiers only.} We tested one strong model (Sonnet 4.6) and one weak model (GPT-5-mini). The optimal capability gap and the generality of our findings across model pairs remain open questions.

\paragraph{Estimated cost metrics.} Our cost analysis (Table~\ref{tab:cost}) is based on estimated token counts from typical prompt sizes and session lengths, not measured API billing data. Exact costs depend on provider pricing, which varies across platforms and changes over time.

\paragraph{Pipeline-specific artifacts.} Some results may be influenced by our specific implementation choices (Copilot CLI, prompt templates, cap values) rather than reflecting fundamental properties of the IC/manager structure.

\paragraph{Non-determinism.} As shown in Section~\ref{sec:analysis}, identical configurations can vary by 10pp on small subsets. Our 200-instance main results are more stable, but the 50-instance ablations should be interpreted as directional rather than precise.

\section{Future Work}
\label{sec:future}

\paragraph{Training for multi-agent roles.} The most direct implication of our analysis is that models should be trained on delegation and scoped execution tasks. Fine-tuning on manager-worker transcripts---where the manager writes task briefs and the worker executes and reports---could develop the skills that current tool-use training does not provide. The question is whether these skills can be added through fine-tuning or require changes to pretraining data.

\paragraph{Capability gap calibration.} Our results show a binary outcome (strong manager helps, weak manager hurts), but the threshold and its dependence on task difficulty remain open questions. Systematic study across multiple model pairs and gap sizes would yield practical guidance for choosing models in production.

\paragraph{Addressing the planning gap.} The manager's inability to plan for unknown unknowns (Section~\ref{sec:analysis}) is the most fundamental structural limitation. Possible mitigations include giving the worker limited autonomy to deviate from the plan when it discovers relevant information, or having the worker flag unexpected findings for the manager to incorporate into a revised plan.

\paragraph{More organizational structures.} Beyond IC/manager, evaluating pair programming (two peers alternate), architecture review (committee evaluates competing plans), and hierarchical management (manager $\to$ tech lead $\to$ workers) would test whether different structures have different training distribution requirements.

\paragraph{Larger-scale evaluation with repetitions.} Extending to SWE-bench Verified (500 instances) with 3 repetitions per configuration, enabling statistical significance testing and variance estimation across the full difficulty spectrum.

\section{Conclusion}
\label{sec:conclusion}

We studied whether an expensive AI model can effectively direct a cheap one to solve software engineering tasks. The answer is \textbf{yes, but conditionally}: the directing relationship requires a genuine capability gap, structured interaction protocols, and careful constraints on what each model can access.

On 200 SWE-bench Lite instances, our \ours{} pipeline (62\%) matches a strong single agent (60\%) while shifting $\sim$90\% of tokens to a free-tier worker model. But the more important contribution is what this structure reveals about current models.

The core finding is that \textbf{current language models are trained as monolithic agents, and splitting them into director/worker roles fights their training distribution}. Directors revert to reading files and forming conclusions instead of delegating. Workers narrate intent instead of executing and reporting. Weak models cannot direct at all---their plans are plausible but wrong, and the structure amplifies rather than compensates for this failure. The pipeline works not by overcoming these limitations but by designing around them: the manager writes text (close to its training), the worker calls tools in scoped contexts (close to its training), and the pipeline code provides the organizational structure that neither model can maintain.

This diagnosis is actionable. It points to specific training gaps---delegation (writing task briefs for others), scoped execution (following instructions without reasoning about the larger problem), and mode switching (knowing when to explore vs.\ when to follow precisely)---that would make multi-agent systems more effective. The guided-then-strict strategy (+10pp) is a prompt engineering workaround for the mode-switching deficit; training models that can regulate this transition internally would reduce the pipeline's dependence on handcrafted phase management.

As models improve, the specific configurations will change. But the underlying insight will persist: splitting reasoning from execution is only productive when each model operates close to its training distribution, and the organizational structure that connects them must be externalized into code rather than entrusted to the models themselves.

All code and evaluation infrastructure are available at \url{https://github.com/Ruil/swe-bench-eval}.


\appendix

\section{Prompt Templates}
\label{app:prompts}

\subsection{Phase 1: Analysis Prompt}

\begin{lstlisting}[language=,basicstyle=\footnotesize\ttfamily]
You are a senior software engineer analyzing a GitHub issue.

## Issue
{problem_statement}

## Your Task
Identify 2-3 specific exploration tasks that a junior
engineer should perform to gather the information needed
to fix this issue.

Each task should be a focused investigation:
- "Find the X method in Y file and report its signature"
- "Check how Z handles the case when W is None"

Output each task on its own line starting with "TASK: "
\end{lstlisting}

\subsection{Phase 4: Implementation Prompt (Round 1, Guided)}

\begin{lstlisting}[language=,basicstyle=\footnotesize\ttfamily]
You are fixing a bug in `{repo}`.

## Issue
{problem_statement}

## Analysis & Plan
A senior engineer has analyzed the issue and written a plan:

{plan}

## Your Task
Implement the fix described above. You have full access
to the repo.

1. Read the relevant file(s) to understand the current
   code structure.
2. Make the changes described in the plan.
3. Run `git diff` to produce your patch.

The plan tells you WHAT to fix and WHERE. Use your
judgment on the exact implementation.
Do NOT modify test files. Keep changes minimal.
\end{lstlisting}

\subsection{Phase 4: Implementation Prompt (Round 2+, Strict)}

\begin{lstlisting}[language=,basicstyle=\footnotesize\ttfamily]
You are fixing a bug in `{repo}`. A senior engineer has
reviewed your previous attempt and written corrected
instructions.

## Corrected Instructions -- follow these EXACTLY
{prior_feedback}

## Original Plan (for context)
{plan}

## Rules
1. Apply ONLY the changes described above.
2. Do NOT modify test files.
3. Output the complete git diff as your final answer.
\end{lstlisting}

\subsection{Manager Failure Review Prompt}

\begin{lstlisting}[language=,basicstyle=\footnotesize\ttfamily]
## Implementation Attempt Failed (round {round_num})

The worker attempted to implement your plan but failed
to produce a valid patch.

Worker output summary:
{worker_output_summary}

Original plan:
{plan}

## Your Task
Analyze why the worker failed and provide revised,
more specific guidance. Consider:
- Was the file path or code location wrong?
- Did the worker need more context about the code?
- Should the approach be simplified?

Write revised instructions the worker can follow.
\end{lstlisting}

\section{Instance-Level Results}
\label{app:instances}

\begin{table}[h]
\centering
\small
\begin{tabular}{@{}ll@{}}
\toprule
\textbf{Solved by \ours{} only} & \textbf{Solved by Strong Direct only} \\
\midrule
\texttt{matplotlib-23299} & \texttt{astropy-14365} \\
\texttt{matplotlib-23563} & \\
\texttt{matplotlib-23987} & \\
\texttt{seaborn-3190} & \\
\texttt{sympy-12171} & \\
\bottomrule
\end{tabular}
\caption{Instances uniquely solved by each approach. 5 unique to \ours{}, only 1 unique to Strong Direct. The manager-worker pipeline subsumes nearly all of the single agent's capability.}
\label{tab:unique}
\end{table}

\end{document}